\documentclass[twoside,slac_one]{revtex4}
\usepackage{graphicx}
\usepackage{fancyhdr}
\usepackage{amsmath} 
\usepackage{bm}
\usepackage{amsxtra}
\usepackage{amssymb}
\usepackage{amsthm}
\usepackage{latexsym}
\usepackage{lscape}

\begin{document}

\title{Status of the Super-B factory Design\footnote{Work supported by the Department of Energy under contract number DE-AC03-76SF00515.}}

%

\author{W.~Wittmer\footnote{wittmer@msu.edu}}
\affiliation{FRIB, Michigan State University, East Lansing, Mi, USA}
\author{K.~Bertsche, A.~Chao, A.~Novokhatski, Y.~Nosochkov,  J.~Seeman, M.~K.~Sullivan, U.~Wienands, S.~Weathersby}
\affiliation{SLAC, Menlo Park, Ca, USA}
\author{A.V.~Bogomyagkov, E.~Levichev, S. Nikitin,P.~Piminov, D.~Shatilov, S.~Sinyatkin, P.~Vobly, I.N.~Okunev}
\affiliation{BINP, Novosibirsk,RU}
\author{B.~Bolzon,L.~Brunetti, A.~Jeremie}
\affiliation{IN2P3-LAPP, Annecy-le-Vieux, F}
\author{M.E.~Biagini, R.~Boni, M.~Boscolo,T.~Demma, A.~Drago, M.~Esposito, S.~Guiducci, S. Liuzzo, M.~Preger, P.~Raimondi, S.~Tomassini, M.~Zobov}
\affiliation{INFN/LNF, Frascati(Roma)), IT}
\author{E.~Paoloni}
\affiliation{INFN \& Universit\`{a} di Pisa, IT}
\author{P.~Fabbricatore, R.~Musenich, S.~Farinon}
\affiliation{INFN \& Universit\`{a} di Genova, IT}
\author{S.~Bettoni}
\affiliation{CERN, Geneva, CH}
\author{F.~Poirier, C.~Rimbault, A.~Variola}
\affiliation{LAL, Orsay, F} 
\author{M.~Baylac, O.~Bourrion, N.~Monseu, C.~Vescovi}
\affiliation{LPSC, Grenoble, F}
\author{A.~Chanc\'{e}}
\affiliation{CEA, Saclay, F}

\begin{abstract}
The SuperB international team continues to optimize the design of an electron-positron collider, 
which will allow the enhanced study of the origins of flavor physics. The project combines the 
best features of a linear collider (high single-collision luminosity) and a storage-ring collider 
(high repetition rate), bringing together all accelerator physics aspects to make a very high 
luminosity of 10$^{36}$ cm$^{-2}$ sec$^{-1}$. This asymmetric-energy collider with a polarized 
electron beam will produce hundreds of millions of B-mesons at the $\Upsilon$(4S) resonance. The present 
design is based on extremely low emittance beams colliding at a large Piwinski angle to allow 
very low $\beta_y^\star$ without the need for ultra short bunches. Use of crab-waist sextupoles 
will enhance the luminosity, suppressing dangerous resonances and allowing for a higher beam-beam
parameter. The project has flexible beam parameters, improved dynamic aperture, and spin-rotators 
in the Low Energy Ring for longitudinal polarization of the electron beam at the Interaction 
Point. Optimized for best colliding-beam performance, the facility may also provide high-brightness 
photon beams for synchrotron radiation applications.

\end{abstract}

\maketitle

\thispagestyle{fancy}


\section{Introduction}
To achieve the design luminosity of $10^{36}$ in operation an ongoing effort of improvement 
is necessary. As part of this process new design options are explored and their effect on
the design parameters studied. No individual option improves every single aspect of the machine and
therefore all solutions have to be carefully compared and the solution with the overall best
performance chosen. Part of this work is the study of unwanted effects, developing methods 
to compensate these and possibly test them in operating facilities. Based on these results 
design specifications and tolerances are developed. We report on the latest results of these 
efforts.


\section{Project Status}
In December 2010 the Italian Ministry for Education, University and Research approved the 
SuperB project~\cite{PressR}. With this decision the project enters a new stage. The current 
conceptual design has been published in \cite{CDR2}. 
The Campus Tor Vergata near Rome, Italy was chosen as site for the facility. This was officially 
announced at the XVII SuperB Workshop and Kick Off Meeting in La Biodola (Isola d'Elba) Italy 
from 28 May 2011 to 02 June 2011.
On October 7$^{th}$ the Laboratory Nicola Cabibbo was offically launched. The new international 
center for fundamental and applied physics is located on the campus of Rome Tor Vergate in
 Italy and will host the SuperB collider.
The next goal is to deliver the engineering 
design report within the next year. Design construction has started.


\section{Latest Design Features}


\subsection{Design Parameters}
All calculations discussed in this paper and \cite{CDR2} are based on the design parameters as 
summarized in table \ref{parameters}.
\begin{table}[ht]
   \centering
   \caption{Super-B parameters.}
   \vspace{-2mm}
   \begin{tabular}{l|c|c}
       \multicolumn{3}{c}{}\\
       \toprule
       \textbf{Parameter}                & \textbf{HER}&\textbf{LER}\\
       \hline
           Luminosity              [$cm^{-2}s^{-1}$]      & \multicolumn{2}{|c}{1.00E36} \\
           Energy                  [GeV]                  & 6.7     & 4.18   \\
           X-Angle (full)          [mrad]                 & \multicolumn{2}{|c}{66} \\
           $\beta_x^\star$         [cm]                   & 2.6     & 3.2    \\
           $\beta_y^\star$         [cm]                   & 0.0253  & 0.0205 \\
           $\epsilon_x$ (with IBS) [nm]                   & 2.07    & 2.37   \\
           $\epsilon_y$            [pm]                   & 5.17    & 5.92   \\
           Bunch length @ I=0      [mm]                   & 4.69    & 4.29   \\
           Beam current            [mA]                   & 1892    & 2447   \\
           Tune shift x                                   & 0.0021  & 0.0033 \\
           Tune shift y                                   & 0.0989  & 0.0955 \\
           $\delta$ @ I$_{max}$    [$\frac{\Delta E}{E}$] & 6.43E-4 & 7.34E-4 \\
           Lifetime                [min]                  & 4.20    & 4.48 \\
     \hline
   \end{tabular}
   \label{parameters}
\vspace{-3mm}
\end{table}
A complete list of parameters can be found at \cite{CDR2}. 


\subsection{IR Layout}
\begin{figure}[htb]                                
    \begin{center}
    \includegraphics[width=10cm]{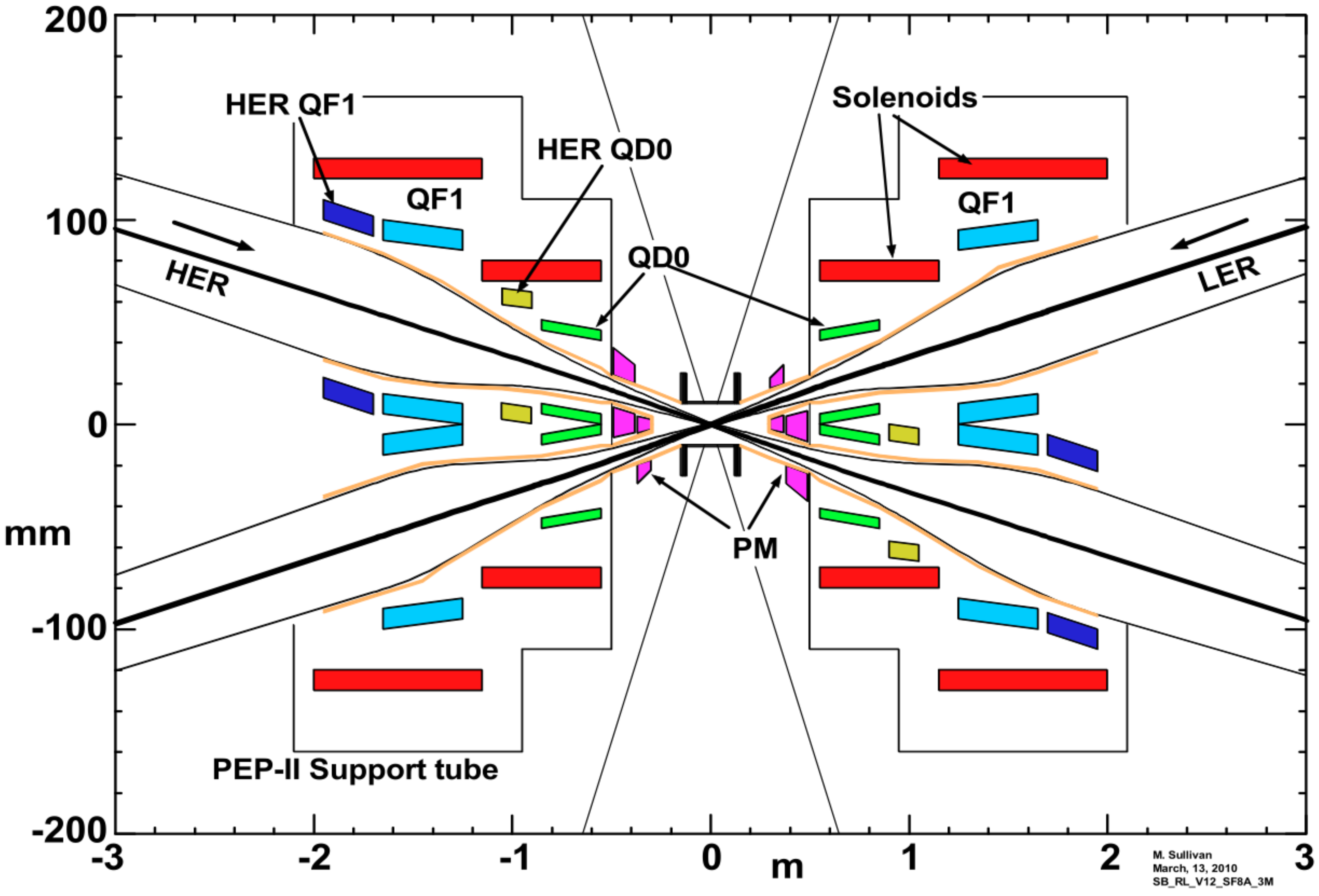} 
    \end{center}
    \caption{Vanadium Permendur ``Russian'' Design.}
    \label{RussianIR}
\end{figure}

There are currently two designs under consideration for the interaction region layout. One bases
its design on vanadium permendur Panofsky quadrupoles for QD0 and QF1 for both rings ("Russian" 
design) as shown in fig.~\ref{RussianIR}.
The second utilizes superconducting air coils for QD0 and QF1 ("Italian" design). In 
addition a portion of the HER QD0 is laid out as vanadium permendur Panofsky quadrupole. 
Both solutions show good behavior for synchrotron radiation backgrounds and beam apertures. 
With respect to the detector solenoid compensation the latter option provides better 
compensation possibilities for both rings as different angles for QD0 and QF1 of 
HER and LER is the optimum solution.   


\subsection{QD0 prototype} 
\begin{figure}[htb]                                
    \begin{center}
    \includegraphics[width=10cm]{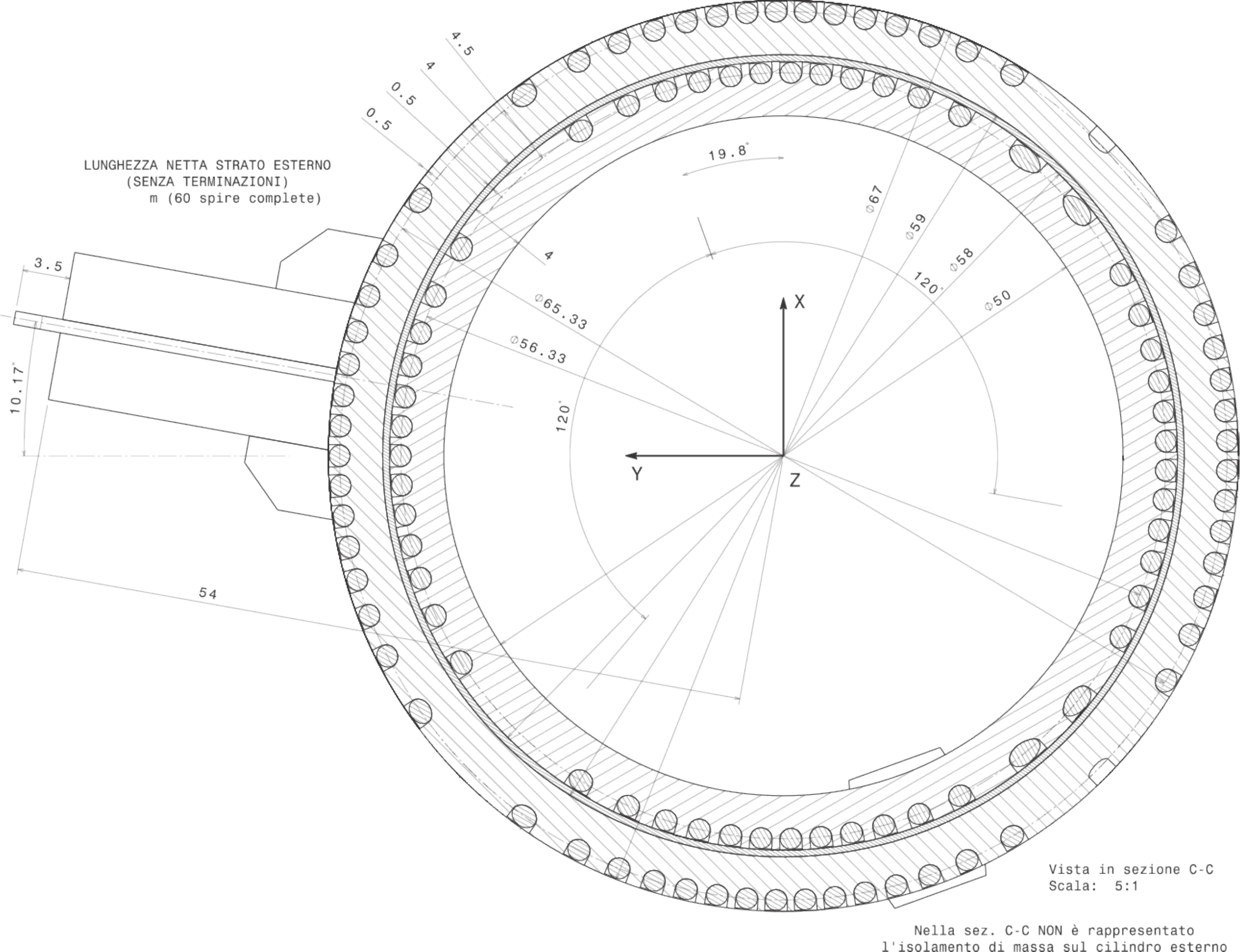} 
    \end{center}
    \caption{Construction drawing for QD0 prototype ''Italien`` style.}
    \label{QD0}
\end{figure}

\begin{figure}[htb]                                
    \begin{center}
    \includegraphics[width=10cm]{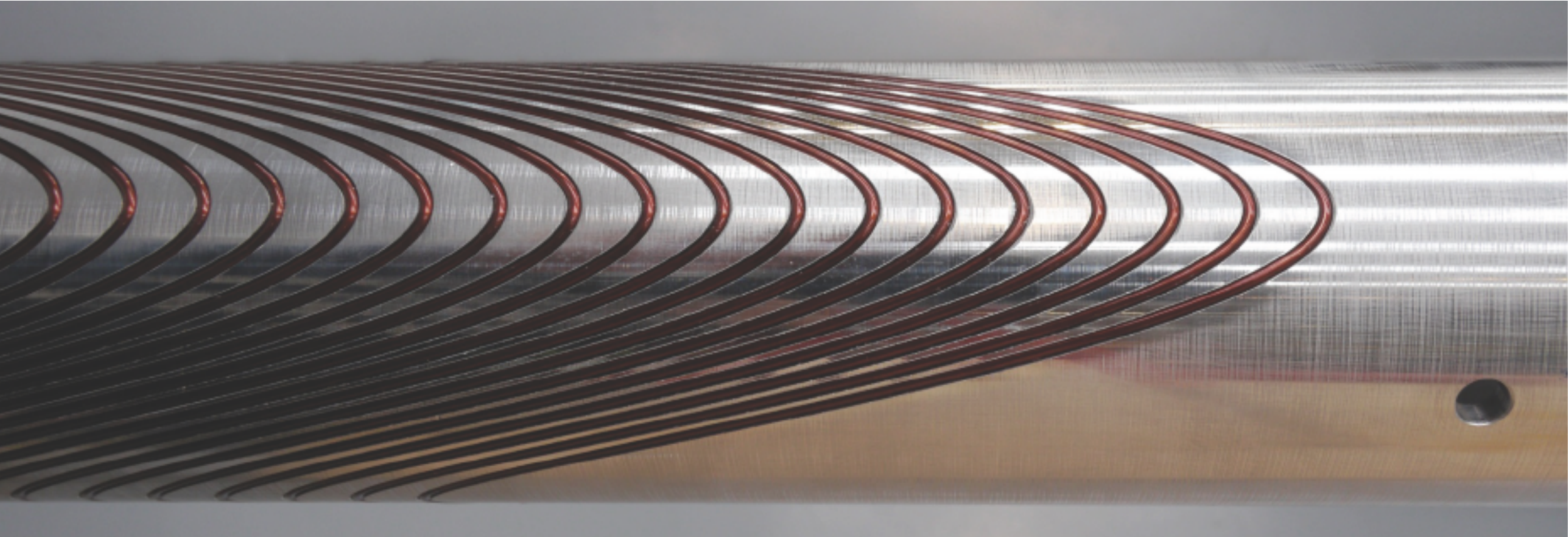} 
    \end{center}
    \caption{Picture showing the aluminum support structure with installed superconducting wire of the ''Italien`` QD0 prototype.}
    \label{QD0_A}
\end{figure}

For the QD0 design a prototype of the superconducting "Italian" design is being built. 
The quadrupole will have a magnetic length of 30 cm and a gradient of 96 T/m and an inner 
bore radius of 2 cm. Critical aspects are the small space available for the SC wire and 
the thermal stabilization material (Cu+Al), and the small margin to quench. The prototype 
will help in determining the maximum gradient achievable at 4.2K and the field quality at 
room temperature. See Fig.~\ref{QD0} for a sketch of the magnet.
The aluminum structure supporting the superconducting wire has been constructed and a test winding installed in it as shown in Fig.~\ref{QD0_A}.


\subsection{Arc Cell Design}
Over the past year two alternative arc cells have been developed. Three different versions 
for the HER arc cell were designed: V12 is the current version, V13 was modified to improve 
chromatic behavior and V14 to reduce the ring circumference.
The major driver were the optimization of emittance, dynamic aperture, reduction of chromaticity
for version 13 and reduction of ring circumference for version 14. The parameters for
these new lattices are summarized in \cite{CalTechSum}. A detailed description for both solutions 
is given in \cite{ArcCellR}.


\subsection{Compensation of Detector Solenoid}
The detector solenoid has a strong 1.5 T field in the IR area, and its tails extend over the 
range of several meters. The coupling of the horizontal and vertical betatron motion needs to 
be corrected in order to preserve the small design beam size at the Interaction Point. Additional 
complications are:
(1) The solenoid is not parallel to either of the two beams due to the crossing angle leading 
to orbit and dispersion perturbations.
(2) The solenoid overlaps the innermost IR permanent quadrupoles causing additional coupling 
effects. 
The proposed correction system 
provides local compensation of 
the solenoid effects independently for each side of the IR. It includes bucking solenoids to 
remove the unwanted longitudinal solenoid field tails and a set of skew quadrupoles, dipole correctors 
and ante-solenoids to cancel all linear perturbations to the optics. The details of the 
correction system design are presented in \cite{Lattice}.


\subsection{Low Emittance Tuning}
The ultra low vertical emittances in SuperB rings need a very careful procedure of emittance
tuning, including the study of the maximum tolerable magnet errors (such as misalignments, 
tilts, BPM errors). 
To minimize the vertical emittance an orbit and dispersion as well as coupling and 
$\beta$-beating free steering routine has been developed. The efficiency of this routine has
been studied for the SuperB lattices with MADX. The results were presented in \cite{LETT}. 
The SuperB final focus (FF) introduces stringent restrictions on alignment of both FF and arcs.
To test the applicability of this method beam studies were conducted in late 2010 at the 
Diamond Light Source. In this test the vertical emittance correction was compared to the 
standard method using LOCO. These studies are
still ongoing and preliminary results were presented in \cite{LETP}. 


\subsection{Vibration Study}
To establish a ``vibration budget'' an ongoing study investigates the effects of on beam 
sensitivity to dynamic misalignments. The major sources for these are expected to be the 
vibration of the IR cryostat. To confirm this the effects of ground motion as originally 
measured at LNF and later on the Tor Vergate site
and the vibration of arc quadrupoles have been compared to the results for both sites.
The results of the Tor Vergate site are favorable compared to the LNF site which contributed
 to the site choice.
For this investigation ground motion measurements at LNF site at Frascati and Tor Vergata, vibration
measurement at the SLD detector at SLAC are combined with theoretical modeling.
To establish the budget a maximum allowed orbit displacement at the interaction point 
reducing the luminosity is defined. The preliminary results are summarized in table 
\ref{vibration}.
\begin{table}[hbt]
   \centering
   \caption{Super-B Proposed Vibration Budget.}
   \begin{tabular}{lcccc}
       &&&&\\
       \toprule
       \textbf{Element}                             & \textbf{RMS}             & \textbf{Transfer}      & \multicolumn{2}{c}{\textbf{RMS differiential}} \\ 
                                                              & \textbf{motion per}  & \textbf{Function}     & \multicolumn{2}{c}{\textbf{displacement at IP}} \\ 
                                                              & \textbf{element}      & \textbf{(RMS sum, }   & \textbf{no}           & \textbf{with} \\ 
                                                              &                                 & \textbf{both rings)}   & \textbf{feedback} & \textbf{feedback} \\ 
       \hline
           Cryostat linear                            & $< 800\,nm$            & $< 0.05$                  & $< 40\,nm$         & $< 4\,nm$ \\
           Cryostat rotation                        & $< 2\,\mu rad$       & $0.02\,\frac{m}{rad}$ & $< 40\,nm$         & $< 4\,nm$ \\
           Final focus quads, excluding IR &  $< 200\,nm$            & $< 0.2$                    & $< 40\,nm$         & $< 4\,nm$ \\
           Arc                                             & $< 200\,nm$             & $< 0.14$                  & $< 30\,nm$         & $< 3\,nm$ \\
           Total (two rings)                        &                                  &                                   & $< 75\,nm$         & $< 7.5\,nm$ \\
       \hline
   \end{tabular}
   \label{vibration}
\end{table}
These results are based on the assumption that the beam feedback achieves a 10 times reduction
of motion at IP. Also the integrated RMS motion is larger than 1 Hertz. By achieving these
values the relative beam motion at the IP will be within 8 nm and the luminosity loss below one
percent. This work has been presented in \cite{vibration}.


\subsection{Effect of Second Order Momentum Compaction}
\begin{figure}[htb]                                
    \begin{center}
    \includegraphics[width=10cm]{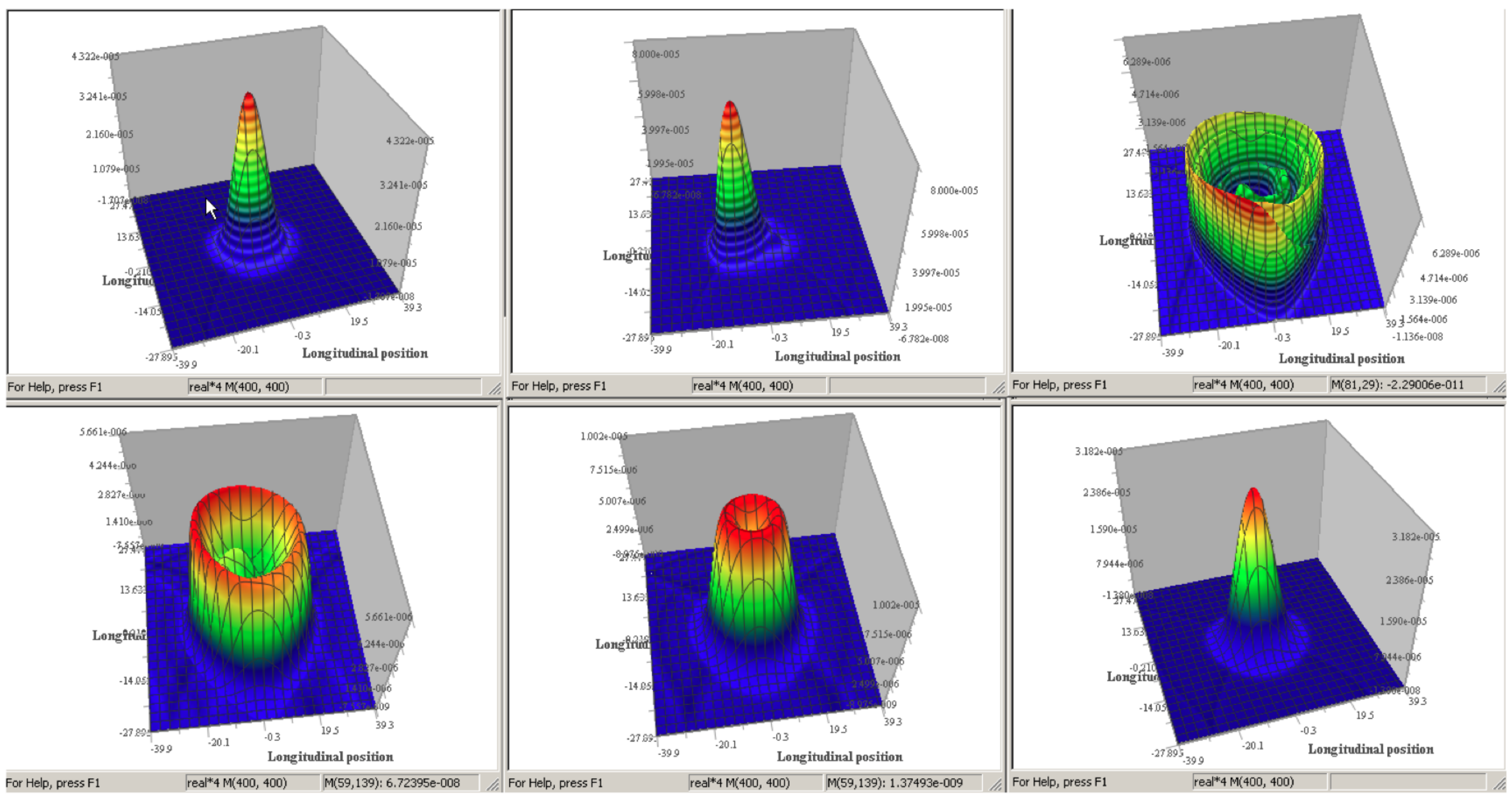} 
    \end{center}
    \caption{Simulation of head tail instability caused by nonlinear momentum compaction and
    wake fields. The instability manifests itself in an saw tooth behavior.}
    \label{f2}
\end{figure}

A longitudinal head-tail instability can appear when the momentum compaction shows a strong 
nonlinear energy dependency~\cite{chao}. The momentum compaction can be developed as a 
function of energy spread $\delta$ as follows: 
$\alpha = \alpha_1 + \alpha_2\, \delta + \alpha_3\, \delta^2$.
This particular instability is affected by the combination of nonlinear momentum compaction 
and beam impedance. To better understand its effect it was studied 
using the code described in \cite{SashaCode}. It showed the behavior of an saw tooth instability
as depicted in fig.~\ref{f2}.
The following preliminary results were obtained: (1) Second order term of momentum compaction 
may play important role in the longitudinal beam dynamics for low compaction machines. 
(2) There is no instability threshold for a positive value of $\frac{\alpha_2}{\alpha_1}$. 
(3) It is necessary to check that the second part of the momentum compaction of the Super-B 
lattice has the right sign, or at least is less than 0.05, if it is positive.
During operation of the DA$\Phi$NE collider this behavior has been observed in the past. 
With this study the effects could be understood and, more important, provide the knowledge 
on how to avoid it. 


\section{Synchrotron Light Option}

The possibility of using the HER and LER storage rings as third generation light source
have been investigated. As an initial step the performance of both rings were compared to 
state of the art facilities in operation, construction and design \cite{LightSource}. 
Both HER and LER compare favorable in this study. As a result the 
synchrotron light option has been made integral part of the Super-B project. A preliminary 
analysis was performed of the ring geometry locating synchrotron light beam lines on the LNF 
site at Frascati. This work is applicable to the Tor Vergata campus. The This is work in progress. 
Adding the functionality of a light source to the project
will impact other decissions.

The design parameters relevant to this study, are shown in Table~\ref{parameter2}. As 
comparison the design parameters from {NSLS II}~\cite{NSLSII} and other state of the
art synchrotron light sources have been added to this table. From these parameters it is 
obvious that synchrotron radiation generated from both HER and LER is comparable to this 
last generation sources. 


\subsection{Benchmark Results}

To quantify the statement above the synchrotron radiation generated from both rings using 
bend magnets and standard undulators have been analytically calculated using the formalism 
described in \cite{XRB}. A MATLAB script
was used to calculate the bessel functions and calculate flux, and brightness. To verify 
this setup a benchmark test was performed with a set of parameters and results calculated 
with an alternative code. As comparison 
the same calculations have been performed for other light sources to benchmark both HER 
and LER. For the comparison state of the art facilities in operation, construction and 
design are used. 


\subsubsection{Bend Magnet}

\begin{figure}[htb]                                
    \begin{center}
    \includegraphics[width=10cm]{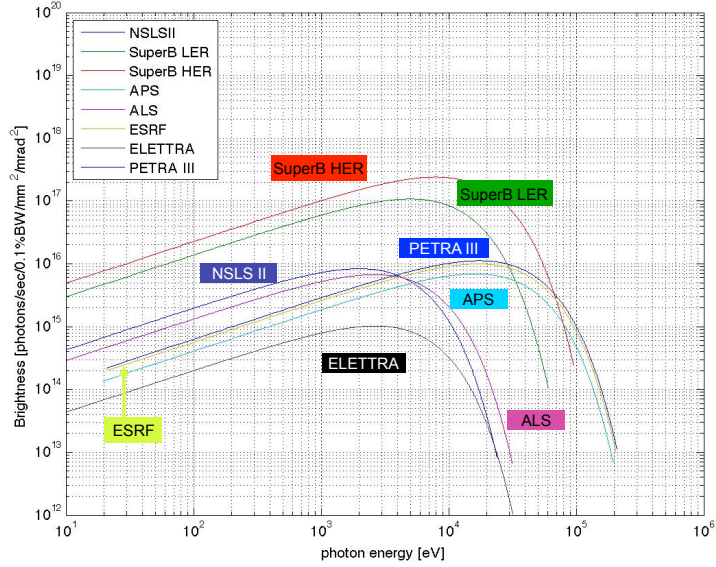} 
    \end{center}
    \caption{Brightness generated from bend magnets as a function of photon energy.}
    \label{f1}
\end{figure}

Bend magnet radiation is parasitically generated in any collider and no change to the 
optics needs to be implemented. Although most third generation light sources are optimized 
for undulator and wiggler beam lines bend magnet sources are part of future designs and 
their performance optimized. The brightness for the different facilities is shown
in Figure \ref{f1}. 
The parameters used for the calculations are those in Table \ref{parameter2}.
The data were extracted from the facilities web pages, CDR's and presentations. 
Future facilities like PEPX are not included here as no data was available for it.
\begin{table*}[htb]
   \centering
   \caption{Parameter Table for Bend Magnet Radiation Calculation.}
   \begin{tabular}{lccccccc}
       &&&&&&&\\
   \toprule
   \textbf{Param.} & \textbf{HER} & \textbf{LER} & \textbf{NSLS II} & \textbf{APS} & \textbf{ESRF} & \textbf{ELETTRA} & \textbf{ALS} \\
   \hline
   E[GeV]               & 6.7  & 4.18 &  3.0  & 7.0   &  6.03  & 2.0  & 1.9  \\ 
   I[mA]                & 1892 & 2447 &  500  & 100   &  200   & 320  & 500  \\ 
   $\rho$[m]            & 69.6 & 26.8 & 24.98 & 38.96 &  23.62 & 5.55 & 4.81 \\ 
   $\epsilon_x$[nm]     & 2.0  & 2.46 &  0.55 & 2.51  &  4.0   & 7.0  & 6.3  \\ 
   $\epsilon_y$[pm]     & 5.0  & 6.15 &  8.0  & 22.6  &  25.0  & 70.0 & 50.0 \\ 
   $\gamma_y$[m$^{-1}$] & 0.33 & 0.54 & 0.05  & 0.101 &  0.10  & 0.50 & 0.74 \\ 
   $\sigma_x$[mm]       & 82.1 & 92.1 & 125   & 81.7  &  77.0  & 139  & 102  \\ 
   $\sigma_y$[mm]       & 8.66 & 9.11 & 13.4  & 27.0  &  29.5  & 28.0 & 8.20 \\ 
   \hline
   \end{tabular}
   \label{parameter2}
\end{table*}
For bend magnets the HER and LER have the brightest photon beam. However the
offset to the other sources is smaller as the source parameters (beam emittance and 
divergence) are less optimized. The results for photon flux is not shown here, since
it is dominated by beam current and the HER and LER will have the highest.


\subsubsection{Undulators}

\begin{figure*}[htb]                                
    \begin{center}
    \includegraphics[width=9cm]{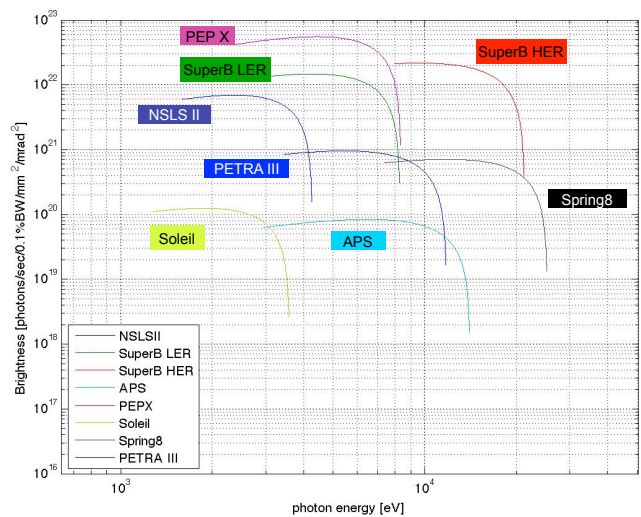} 
    \end{center}
    \caption{Brightness as a function of photon energy for different reference
    undulator radiation calculated for benchmarking different facilities.}
    \label{f22}
\end{figure*}

The figure of merit for third generation light sources is the radiation generated from 
insertion devices. For this purpose the undulator radiation as documented on the same
sources quoted above have been used for facilities in operation and construction. For
better comparison the same device type as at NSLS II was chosen for the Super-B 
HER and LER. Source point and device parameters are summarized in Table \ref{parameter3}.
\begin{table*}[hbt]
   \centering
   \caption{Parameter Table for Undulator Radiation Calculation.}
   \begin{tabular}{lcccccccc}
       &&&&\\
   \toprule
   \textbf{Param.}    & \textbf{HER}   & \textbf{LER}   & \textbf{NSLS II} & \textbf{APS} & \textbf{PEPX}  & \textbf{Soleil} & \textbf{Spring8} &  \textbf{Petra III} \\
   \textbf{Undulator} & \textbf{IVU20} & \textbf{IVU20} & \textbf{IVU20}   & \textbf{U33} & \textbf{IVU23} & \textbf{U20}    & \textbf{U24}     & \textbf{U29} \\
   \hline
   E[GeV]             & 6.7  & 4.18 & 3.0  & 7.0  & 4.5  & 2.75 & 8.0  & 6.0  \\ 
   I[mA]              & 1892 & 2447 & 500  & 100  & 1500 & 500  & 100  & 100  \\ 
   $\sigma_x$[$\mu$m] & 82.1 & 92.1 & 125  & 81.7 & 22.2 & 3880 & 286  & 140  \\ 
   $\sigma_y$[$\mu$m] & 8.66 & 9.11 & 13.4 & 27.0 & 7.00 & 8.08 & 6.00 & 5.60 \\ 
   $\sigma_x'$[mrad]  & 33.3 & 37.0 & 16.5 & 11.8 & 7.40 & 14.5 & 11.0 & 7.9  \\ 
   $\sigma_y'$[mrad]  & 2.1  & 2.7  & 2.7  & 3.3  & 1.2  & 4.6  & 1.0  & 4.1  \\ 
   N[1]               & 148  & 148  & 148  & 72   & 150  & 90   & 186  & 172  \\ 
   $\lambda_u$[mm]    & 20   & 20   & 20   & 33   & 23   & 20   & 24   & 29   \\ 
   K$_{max}$[1]       & 1.83 & 1.83 & 1.83 & 2.75 & 2.26 & 1.0  & 2.21 & 2.2  \\ 
   K$_{min}$[1]       & 0.1  & 0.1  & 0.1  & 0.1  & 0.1  & 0.1  & 0.1  & 0.1  \\ 
   \hline
   \end{tabular}
   \label{parameter3}
\end{table*}
The brightness for the undulator radiation is depicted in Figure \ref{f22}.
PEPX generates the highest brightness followed by HER and LER. This is a result of 
the high beam currents in these rings. A detailed study of the difference of source
point parameter and its effect on the brightness between PEPX and the Super-B rings
by considering the photon beam diffraction limit showed that the higher brightness
in PEPX is generated by its lower horizontal emittance.


\subsubsection{Source Parameter Optimization}

\begin{figure}[htb]                                
    \begin{center}
    \includegraphics[width=9cm]{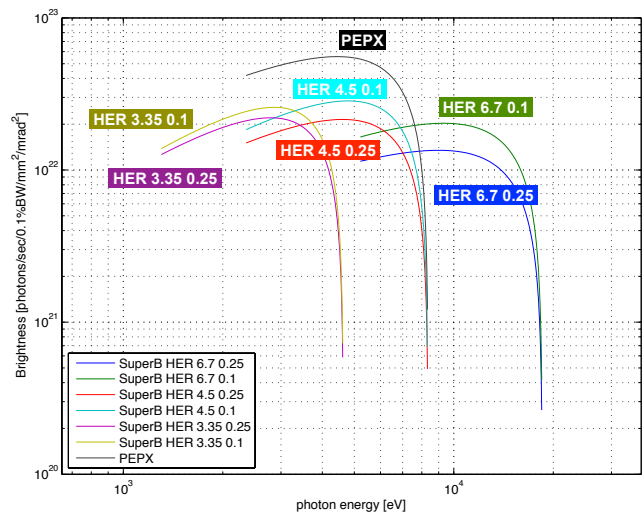}
    \end{center}
    \caption{Brightness as a function of photon energy for different energies
    and coupling values of the HER in comparison to PEPX baseline design.}
    \label{f3}
\end{figure}

A brief study to optimize the brightness in the HER was conducted, the result is 
plotted in Figure~\ref{f3}. By reducing the HER energy and beam coupling (from design 
0.25\% to 0.1\%) the beam emittance is reduced. This gain is unfortunately mostly 
canceled by the change of the diffraction limit as the 
calculation show, since to achieve the same level of brightness as in PEPX the 
horizontal emittance would have to be significantly reduced.
%


\subsection{Beam Line Layout}

When using the Super-B rings as a synchrotron light source one has to plan
for additional space for the photon beam lines and user space. This is
significant as collider facilities are usually located underground for
radiation protection purposes whereas third generation light sources are 
surface installations.  


\subsubsection{LNF Site at Frascati Rome}

\begin{figure}[htb]                                
    \begin{center}
    \includegraphics[width=10cm]{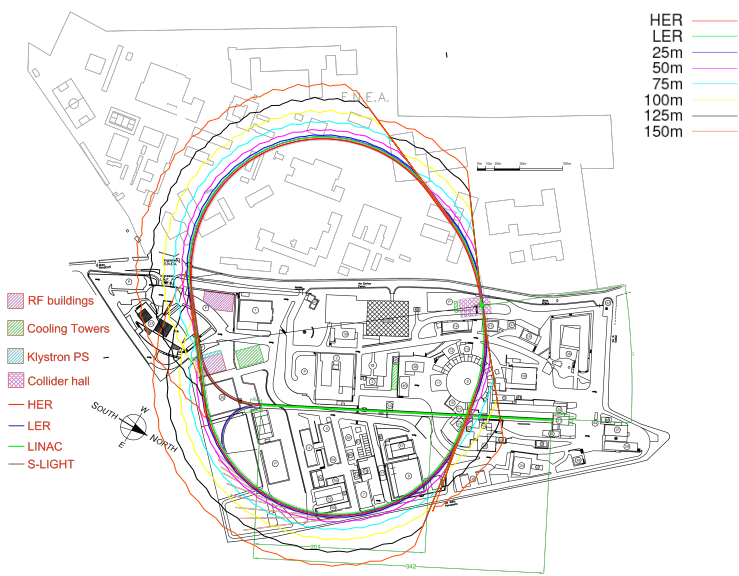} 
    \end{center}
    \caption{All possible bend magnet photon beam line location with different
     contour lines showing the distance to the source point.}
    \label{f4}
\end{figure}

\begin{figure*}[ht]                                
    \begin{center}
    \includegraphics[width=10cm]{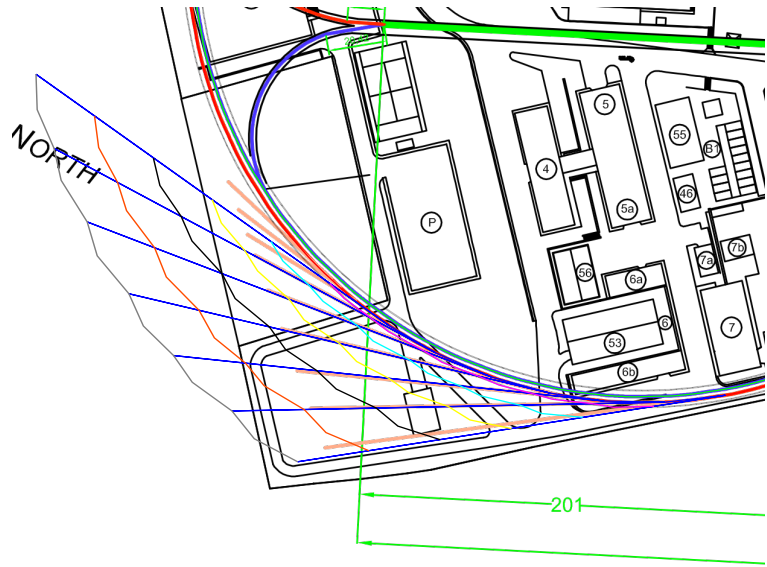} 
    \end{center}
    \caption{Area of possible HER beam lines which would be located within
    LNF boundaries. The contour line legend of fig.\ref{f4} applies.}
    \label{f5}
\end{figure*}

During the study the LNF site near Rome was the primary choice. This site
demands an underground installation as the surface area is already mostly 
occupied. In addition the ring dimensions partially exceed the LNF boundaries. 
This investigation was aimed to understand where and with what length 
photon beam lines could be placed within the LNF boundary. An overview 
is shown in Figure \ref{f4}. It shows the planned Super-B tunnels with 
injector. Added to these are the contour lines with distances from the 
source point in 25 meter steps to a maximum of 150 meters. Only bend magnet
source points from the outer ring were used. the upper portion of beam 
lines originate from the LER the lower from the HER.
Figure \ref{f5} is a magnification of the lower left area. As the picture
indicates a set of HER beam lines with different length could be located 
there completely within LNF boundaries. The same contour line legend of 
Fig.\ref{f4} applies.


\subsubsection{Campus Tor Vergata Rome}

\begin{figure*}[ht]                                
    \begin{center}
    \includegraphics[width=10cm]{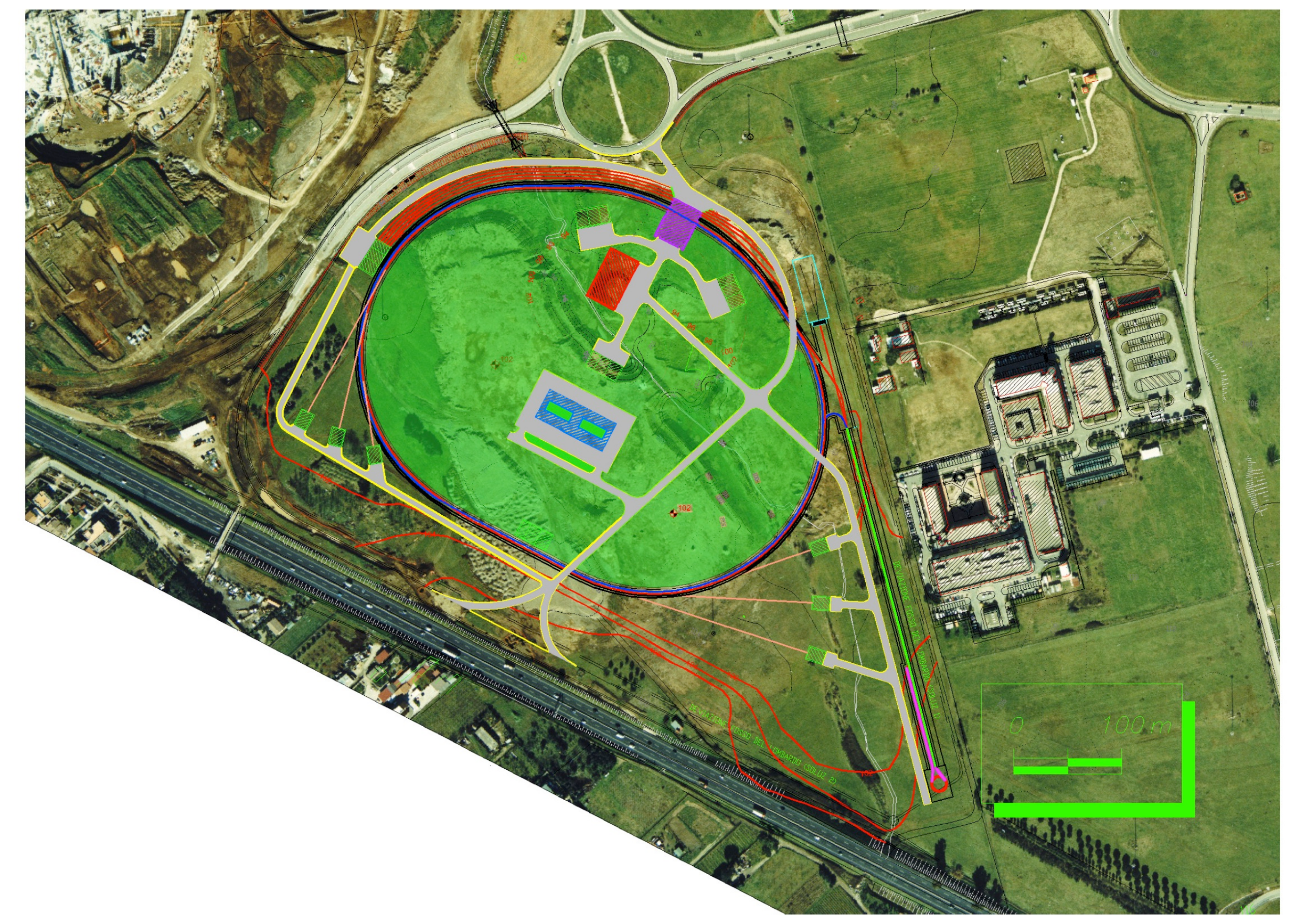} 
    \end{center}
    \caption{Design option of the SuperB collider at the Campus Tor Vergata near Rome.}
    \label{TORVERGATA}
\end{figure*}

With the decission in favour for the Tor Vergate site the space constraints discussed in the previous section
are relaxed. This site is located about 4 kilometers from the original area of LNF. The preliminary results of
the ground motion measurements show less activity at this site. A first scetch of the accelerator design at 
Tor Vergata as originaly planned for the LNF site is shown in Fig.~\ref{TORVERGATA}.


\section{CONCLUSIONS}
With the approval by the Italian government the project is moving from the conceptual to the
technical design phase. With the establishing of the Laboratory Nicola Cabibbo at the Campus 
Tor Vergata as the future site of the SuperB facility a large step has been taken towards the realization.
The collaboration with other funding agencies, through MOU, is in its final phase. Construction plans will follow soon.


\end{document}